# UV to far-IR reflectance spectra of carbonaceous chondrites. I. Implications for remote characterization of dark primitive asteroids targeted by sample-return missions

Josep M. Trigo-Rodriguez[1],
Carles E. Moyano-Cambero[1],
Jordi Llorca[2],
Sonia Fornasier[3],
Maria A. Barucci[3],
Irina Belskaya[3],
Zita Martins[4],
Andy S. Rivkin[5],
Elisabetta Dotto[6]
José M. Madiedo[7, 8], and
Jacinto Alonso-Azcárate[9]

[1] Institute of Space Sciences (CSIC-IEEC), Campus UAB, Facultat de Ciències, Torre C-5, parells, 2ª planta, 08193 Bellaterra (Barcelona), SPAIN
[2] Institut de Tècniques Energètiques i Centre de Recerca en Nanoenginyeria. Universitat Politècnica de Catalunya, Diagonal 647, ETSEIB. Barcelona, Spain.
[3] LESIA, Observatoire de Paris, CNRS, UPMC Univ Paris 06, Univ. Paris-Diderot, 5 Place Jules Janssen, 92195 Meudon Principal Cedex, France
[4] Department of Earth Science and Engineering, Imperial College London, London SW7 2AZ, UK.
[5] John Hopkins University Applied Physics Laboratory, Laurel, MD, USA.
[6] INAF-Osservatorio Astronomico di Roma, Rome, Italy.
[7] Facultad de Ciencias Experimentales, Universidad de Huelva, Huelva, Spain.
[8] Departamento de Física Atómica, Molecular y Nuclear. Facultad de Física, 41012 Sevilla, Spain.
[9] Universidad de Castilla-La Mancha, Campus Fábrica de Armas, 45071 Toledo, Spain.

Abstract: We analyze here a wide sample of carbonaceous chondrites from historic falls (e.g. Allende, Cold Bokkeveld, Kainsaz, Leoville, Murchison, Murray, Orgueil, and Tagish Lake), and from NASA Antarctic collection in order to get clues on the role of aqueous alteration in promoting the reflectance spectra diversity evidenced in the most primitive chondrite groups. We particularly focus in the identification of spectral features and behavior that can be used to remotely identify primitive carbonaceous asteroids. The selected meteorite specimens are a sample large enough to exemplify how laboratory reflectance spectra of rare groups of carbonaceous chondrites exhibit distinctive features that can be used to remotely characterize the spectra of primitive asteroids. Our spectra cover the full electromagnetic spectrum from 0.2 to 25 µm by using two spectrometers. First one is a UV-NIR spectrometer that covers the 0.2 to 2 µm window, while the second one is an Attenuated Total Reflectance IR spectrometer covering the 2 to 25 µm window. In particular, laboratory analyses in the UV-NIR window allow obtaining absolute reflectance by using standardized measurement procedures. We obtained reflectance spectra of specimens belonging to the CI, CM, CV, CR, CO, CK, CH, R, and CB groups of carbonaceous chondrites plus some ungrouped ones, and allows identifying characteristic features and bands for each class, plus getting clues on the influence of parent body aqueous alteration. These laboratory spectra can be compared with the remote spectra of asteroids, but the effects of terrestrial alteration forming (oxy)hydroxides need to be considered.

1. Introduction

In the past few decades, parallel studies of the composition, petrology and reflectance spectra of meteorites have allowed the identification of achondrite parent bodies and clues to the physical properties of asteroids to be inferred (Johnson & Fanale, 1973), offering an perspective on the complex collisional and dynamical history and evolution of asteroids (Gaffey 1976). A clear example of the relevance of obtaining reflectance spectra of asteroids was the successful identification of the Howardite, Eucrite, and Diogenite (HED) suite of achondrites as free-delivered samples from the asteroid 4 Vesta. Such significant success was achieved due to the characteristic olivine and pyroxene absorption bands (for a review see e.g. Pieters & McFadden 1994). Despite such progress, the identification of other asteroidal classes has remained controversial due to several factors: First, the intrinsic spectral diversity of some asteroid surfaces due to compositional differences across their surfaces. Second, changes in reflectance spectra produced by space weathering, caused by cosmic rays and meteoroid bombardment and resulting in slope and band depth changes that are still being explored (Gaffey 2010; Clark et al. 2002). Third, and sometimes overlooked, asteroid spectra are usually taken in non-ideal observing circumstances, subjected to geometrical and phase angle effects that make absolute calibration not always feasible (Sanchez et al. 2012). Some of the above mentioned problems are direct consequence of the intrinsic heterogeneity in the surface of undifferentiated bodies that can exhibit averaged spectra significantly different from the one exhibited by a specific meteorite coming from them. In fact, recent evidence suggests that most chondrites are complex mixtures of different materials, or breccias, produced by the progressive impact gardening of asteroidal surfaces by projectiles of very different composition (Bischoff et al. 2006). Due to all these circumstances, meteorite spectra obtained in laboratories might differ significantly from the spectral appearance of their respective parent asteroids, increasing the magnitude of this puzzle (see e.g. Burbine 2002).

Significant progress has been made since pioneering work concerning the characterization of asteroids in distinctive classes according to their spectra (Tholen 1984, 1989; Barucci et al. 1987; Bus and Binzel 2002; Clark et al. 2011; Ostrowski et al. 2011). Guided by accurate meteorite spectra obtained in terrestrial laboratories, direct comparisons were made, and potential matches identified. From the identified groups of chondrites and achondrites, the spectroscopists were then able to find distinctive features, bands, and slopes that help them to remotely identify the presence of certain minerals, and also define asteroid classes. Significant differences between asteroid classes and other asteroid classes or likely meteorite analogs were found and they were later explained as consequence of different processes: space weathering, parent body aqueous alteration, different regolith sizes, etc.

Some dark and primitive asteroids are future targets of space based missions like e.g. OSIRIS-REx or Marco Polo-R (Lauretta et al., 2012; Barucci et al., 2012a). However, their successful characterization, and association with meteorite analogs remains in most cases elusive. Asteroids of the C class (or the C complex in the Bus-DeMeo taxonomy), which are the targets of the next round of sample return missions, are mostly characterized by their low albedos, and flat, mostly featureless spectra. The C class was first temptatively associated with the CI and CM groups of carbonaceous chondrites (hereafter CCs) (Vilas & Gaffey, 1989). Later on, it was found the presence of water absorption features in about half of the C complex asteroids, particularly at 0.7 and 3 µm (Lebofsky, 1978, 1980; Vilas & Gaffey, 1989; Jones et al. 1990, Fornasier et al., 1999). The CO and CV chondrites were associated with K asteroids by Bell (1988), and the CR group was linked to C asteroids by Hiroi et al. (1996). Finally, the rare CH chondrite group was linked to C or M asteroids by Burbine et al. (2002). Despite of this, most of these links are temptatively assigned so any further progress in characterizing the asteroids that are sources of meteorites is particularly relevant.

Probably the most poorly known asteroid classes are those associated with undifferentiated carbon-rich asteroids that are linked with CCs (see Table 1). These objects are very primitive and are linked with highly porous and fragile carbonaceous chondrites (Consolmagno and Britt 1998; Britt & Consolmagno 2000) that rarely survive atmospheric entry (Trigo-Rodríguez & Blum 2009). Not all CC meteorites have reflectance properties systematically studied, but recent comprehensive studies exist dealing with powders of the most abundant ones: CI, CM and CR (Cloutis et al. 2011a,b; 2012a,b,c,d,e,f). In this paper we analyze carbonaceous chondrites from Antarctica with the goal of identifying diagnostic absorption features in both common and rare CC groups. Using two different spectrometers we have measured their reflective properties in a wide wavelength range: 0.2 to 25 µm. Some of the patterns and features found in our spectra have some promise to allow us to distinguish between different chondrite groups. Finally, as a byproduct of our current study we have found a plausible link between the CH chondrites and the asteroid 21 Lutetia recently visited by Rosetta spacecraft.

Table 1.

2. Analytical procedure and data reduction.

The meteorites studied here have been carefully selected from the NASA Antarctic collection. As noted above, the spectra presented here cover the full wavelength range from 0.2 to 30 µm by using two spectrometers. The first one is a Shimadzu UV3600 UV-Vis-NIR spectrometer that covers the 0.2 to 2.6 µm window, while the second one is a Smart Orbit ATR (Attenuated Total Reflectance) IR spectrometer that covers the 2.5 to 40 µm window. Both spectrometers allow us to infer the reflectivity of the meteorites in a very wide range, but with great detail in crucial spectral windows for comparison with the remote spectra of asteroids taken from ground or space-based telescopes.

In reference with the meteorite samples, to obtain the UV-NIR spectra we used mostly thin (~30 µm), but also some thick sections (with a thickness of a few mm) were measured (e.g. Allende or Kainsaz). To avoid unwished contributions from the surroundings, all samples were having larger surface than the beam slot used by the spectrograph. On the other hand, to make sure that the thin sections were not too thin, and therefore were not allowing some of the incident light being transmitted through the section and be lost, we performed an analysis of some meteorites in thin and thick sections obtaining similar results.

The Shimadzu UV3600 UV-Vis-NIR spectrometer allows the measurement of the transmission, absorbance and reflectance spectra of powder, solid, or liquid samples. The standard stage for the spectrometer is an Integrating Sphere (ISR) with a working range of 185 to 2,600 nm. The spectrometer uses multiple lamps, detectors and diffraction gratings to work over a wide range of wavelengths. The light originates at one of two lamps, passes through a variable slit, is filtered to select the desired wavelength with a diffraction grating, and is then split into two alternating but identical beams with a chopper. The sample beam interacts with the sample and is routed to one of two or three detectors (depending on the sample stage). The reference beam interacts with the reference material and then goes to the same detector. The inside of the ISR is coated with a duraflect reflecting polymer. For calibration of the detector a standard baseline was created using a conventional $BaSO_4$ substrate that provided a ~100% reflectance signal better than $1\sigma$ all along the 200-2,000 nm range. Over that range the $BaSO_4$ reflectivity is dropping off so the measurements here must be taken with caution and are not considered in this work. The sampled area during the measurements corresponds to a slot of $2\times1$ cm$^2$ that was larger than the common size of the analyzed samples. A comparative view with these remote spectra is possible, but the major mineral absorption bands obtained by our Shimadzu are almost saturated; therefore we cannot reliably extract band positions so we restrict ourselves to identify the more clear features in UV-NIR.

The received chips of each meteorite sample were ground using an agate mortar to obtain the IR spectra. Only a few minutes elapsed between grinding the samples to powder and placing them in between a diamond detector of the Smart Orbit ATR (Attenuated Total Reflectance) IR spectrometer. The diamond ATR detector is ideal for the analysis of hard materials of very different nature, but particularly meteorites because it is inert and extremely strong to deal with tough chondritic materials (e.g. metals or refractory inclusions). At the same time, this diamond-based detector has a wide spectral range and good depth of penetration, which makes it a good choice for meteoritic samples. In this work we focus in the overall description of the spectra and the identification of the main absorption lines in the wavelength range from 2 to 25 μm. Over 30 μm we noticed a higher data scattering due to laboratory conditions, and therefore we avoid a discussion here (Trigo-Rodríguez et al., 2012). In general our ATR IR spectrometer provides high resolution internal reflection spectra of meteorite powders following standard procedures. We remark that these results are similar in their general trends to the obtained by other techniques (e.g. IR micro-spectroscopy) (Beck et al., 2010)

3. Results.

In our study, we have included samples of relatively well known and common carbonaceous chondrite groups (e.g. CMs, COs and CVs), that can be directly compared with much rare specimens (e.g. those belonging to the CH, CK, CR and R groups) and also to a few ungrouped carbonaceous chondrites. Table 2 compiles all analyzed meteorites. We have found evidence for a continuum in the reflectivity of chondrite groups, from very low reflectance groups to high reflectance ones represented by CHs and CBs. It is remarkable that the last four cited groups have distinctive reflectance properties, particularly higher reflectivity than those found in other chondrite groups. Our measurements in PCA91467 thin section indicate that this CH chondrite is rich in 10 to 200 micron-sized metal grains, also characteristic of the CH group of carbonaceous chondrites that might raise somehow the albedo (Campbell and Humayun, 2004). MAC 02675 is a CB that exhibits few chondrules embedded in the Fe-Ni alloy structure (see e.g. Weissberg et al., 2006).

Table 2

The main mineralogical properties, like e.g. the endmember composition of olivine and pyroxene, and a list of secondary minerals that can play a role in the spectral features observed in the spectra are compiled in Table 3.

### 3.1. Carbonaceous chondrites spectra from UV to NIR.

The reflectance spectra of different groups of carbonaceous chondrites reflect a large compositional diversity. Spectral diversity is seen among the presumed carbonaceous chondrite parent asteroids, reflecting the diversity in mineral components seen in the meteorites (DeMeo et al., 2010; de León et al., 2012). Most data in the literature covers spectral regions 0.4-1.7 µm (for instance, the SMASS, S3OS2, and SMASSIR surveys), although the *RELAB* comprehensive database and current commonly-used asteroid instruments typically reach 2.5 µm. Our data extends this window to wavelengths as short as 0.2 µm and to 2.6 microns as a long-wavelength limit. This is particularly interesting as absolute reflectivity measurements in those windows are rare. We have found e.g. that below 0.4 µm the samples reflectivity tends to converge into a 5 to 10% range. Longward of 2.0 µm the reflectivity diversifies among 15 to 35% and this seems to be direct consequence of the non-ideal reflectivity of our $BaSO_4$ standard for those wavelengths. Consequently, we decided to cut the plotted data at 2 µm.

Polished sections of the selected meteorites were measured using the UV-Vis-nIR spectrometer. The reflectance spectra were obtained at least two times to get an average value for the reflectance in each wavelength, and no significant differences were found by rotating the sections. The slot size was chosen to almost cover the size of the section, usually about $1cm^2$, just following the same procedure explained in previous studies (Trigo-Rodríguez et al., 2011, 2012).

However, we are dealing with polished samples, so we would like to know what are the main differences with the reflectance spectra obtained from grained powders of some of these meteorites by Cloutis et al. (2011a,b, 2012a,b,c,d,e,f). We choose to separate the spectra by chondrite groups to test the homogeneity of reflectance properties for meteorites presumably coming from the same parent body. It has been suggested, from the major element differences among the chondrite groups that each one comes from a different formation location in the protoplanetary disk and probably accreted in a different parent body (Wasson, 1985). In any case, as meteorites arriving to Earth have presumably experienced quite complex and peculiar evolutionary histories, important compositional and mineralogical properties among the different specimens can exist. The later statement is reinforced by the fact that recovered meteorites are small samples of their parent asteroids, probably representative of local processes in their evolutionary paths.

Quite distinctive reflectance spectra were obtained for the different chondrite groups, and to make easy an overall discussion we have separated the spectra in different graphs (see Fig. 1). Reflectance spectra of CM and CK chondrites are plotted in Fig. 1a. Spectra of CR and R chondrites are plotted in Fig. 1b. Finally, CO and CV are plotted in Fig. 1c. To have a more general view we have compiled in Fig. 1d the reflectivity of all samples including the CH and CB analyzed specimens. We will discuss the main features in section 4.1.

### 3.2. Carbonaceous chondrites spectra from NIR to far-IR.

Significant differences in the reflectance spectra of CCs (differing spectral slopes, presence or absence of absorption bands, overall reflectance levels, and so on) suggest compositional diversity. Sometimes the IR spectra are plotted as function of emissivity, but we decided to use the direct transformation provided by our Smart Orbit ATR spectrometer to plot them directly as relative reflectance (Fig. 4-7). The spectral window goes from 2.5 to 45.5 µm with a resolution of about 0.1 µm. The grinding procedure of the meteorite samples was always similar, but depending on the toughness and specific mineralogy of each meteorite sample we noticed that the resulting averaged size of grains was slightly different. Performing some direct measurements of the size of the grains we concluded that was about 100 ± 50 µm. The granulometric studies that we made for example for Allende indicate that a significant population of the grained materials produce powders in the range of few hundred µm that might have direct implications in the inferred positions of the Christiansen feature and the Reststrahlen bands in each chondrite group (see e.g. Mustard and Hays, 1997). Depending on the matrix abundance and degree of alteration of mafic silicates forming chondrules, we observed that the grinding was more efficient for altered chondrites.

CC groups exhibit different components and distinctive abundance ratios, so initially we expected significant differences among the studied groups. Chondrules are the most abundant ingredients, but they vary in average size and proportions, like also occurs for the Ca-Al rich inclusions (CAIs) and other refractory oxides (Rubin et al., 1989). Metal grain abundances are also highly variable from being almost absent to ubiquitous depending on the chondrite group, and their presence inside the chondrules or in the matrix has direct implications for the reflectance. For instance, the abundance of metal grains is probably the reason for the uniformly red-sloped spectra of CRs. Further complicating the interpretation, some water rich groups (e.g. the CMs) exhibit secondary minerals formed by precipitation from a fluid (Trigo-Rodríguez et al., 2006; Rubin et al. 2007). To exemplify the complexity and heterogeneity of aqueous alteration, we know that the mobilization of soluble elements by water was a local process in CMs (Trigo-Rodríguez et al., 2006). It has been suggested that the availability of heat and/or water could have been depth dependent in the parent asteroid (Rubin et al. 2007). Consequently, the reflectance spectra of CM chondrites exhibit different degrees of aqueous alteration and the extent of aqueous products is highly variable.

Each chondrite group exhibits distinctive absorption bands that support the notion that secondary variations in mineralogy influence the meteorite reflectance spectra. On the other hand, among meteorite members of each group exists important differences in water adsorption. Most of that water was probably incorporated in the parent asteroids but it is also known that carbonaceous chondrites incorporate terrestrial water into their mineral structure after their falls (Fuchs et al., 1973).

By using the ATR technique it is possible to identify the main absorption hydroxyl bands corresponding to clay minerals, and also several functional groups of organic compounds present in the C-rich matrix of carbonaceous chondrites (Table 4). Then, it seems that this technique has an important potential to identify the presence of such components in CCs. The following graphs exemplify how the different groups can be identified obtaining such ATR IR spectra.

Table 4.

In Fig. 2 are plotted ten CM chondrites together with Murchison that, due to its availability, is usually considered as a proxy. Aqueous alteration seems to be directly associated with meteorite reflectance. Note for example that the less reflective LEW 87148, QUE 97990, and QUE 99355 are also exhibiting the deepest water absorption bands, then suggesting that reflectivity is also affected by aqueous processing.

As an example of the applicability of these spectra to characterize asteroid surfaces, we note that the 3 μm OH band was identified in the IR spectrum of the near-Earth objects (NEOs) 1996 FG3 that is the backup target of Marco Polo-R mission (see e.g. Fig. 1 in Rivkin et al., 2012) and it has been also identified in other NEO called: 1992 UY4. It is also seen in most of the C-complex objects in the main asteroid belt (Jones et al. 1990, Rivkin et al. 2002, 2003).

In Fig. 3 we plotted four CO chondrites, two CV chondrites, and again CI Orgueil for general comparison. The differences in the location of the main absorption bands compared with Orgueil (but also with the CMs) are evident in such figure, and consequence of different bulk mineralogy (see e.g. Cloutis et al., 2011a, 2012c, 2012d). In Fig. 4 we compiled the rest of spectra belonging to CB, CH, CK and R chondrites. The R3 PRE 95404 and the CK4 ALH 85002 have absorption bands in similar locations than CO and CV chondrites.

Surprisingly we found a spectral similitude (peaks in similar locations) between a preliminarily classified CM QUE 99038 with the reflectance spectra of CVs. We suggested Alan Rubin that QUE 99038 was misclassified or having an anomalous composition, so he reported to the Meteoritical Society a reclassification as CV3 chondrite (Rubin, pers. comm.). This assignation is already established in the Meteoritical Bulletin online database. The affinity with CV3 is also suggested by Figure 5 where we can directly compare QUE 99038 reflectance spectrum with the spectrum of CV3 chondrite Allende, the CK4 ALH 85002, and the R3 PRE 95404. In particular, QUE 99038 features and bands are much closer to those exhibited by the R3 chondrite, even in the position of the olivine peak that occurs at about 12.05 μm, instead of 12.12 μm for CV3 Allende. We also show CM2 Murchison to exemplify that this method allows identifying misclassified chondrites. Cloutis et al. (2012f) remarked that QUE 99038 has a reflectance spectrum similar to the CO3 chondrite ALH77003 in the 1 μm window, but clearly differing in the 2 μm region in which QUE 99038 exhibits a dominant olivine absorption band.

Fig. 2-5

4. Discussion.

4.1. General reflectance patterns for CC groups.

In next subsections we will discuss the main reflectance patterns observed in primitive CCs. Each spectral window exhibits distinctive spectral features that can be used to characterize the nature of a primitive asteroid, the main goal after all.

4.1.1. General patterns seen in the UV to NIR spectral window.

CCs exhibit a wide diversity in reflectance that is a direct consequence of their original and current mineralogy. The similarities in UV to NIR spectra that many groups share complicate efforts at interpretations that use only that spectral range. For example, the CM and CK chondrite groups exhibit absolute reflectivity values between 5 and 10% (see Figure 1a). The main characteristics of the CM chondrite spectra are a modestly blue-, flat or red-sloped profiles, the presence of an ubiquitous absorption band in the 700 nm region, and a broader region of absorption between ~900 and 1,200 nm attributable to olivine and phyllosilicates (Cloutis et al. 2011b). Our results also confirm that the respective depths of the previously cited absorption bands vary largely so they should be consequence of different degree of aqueous alteration or, in other words, of the presence of serpentine-rich silicates in their mineral structure (Cloutis et al., 2011b). Table 3 compiles the main mineralogical characteristics of the analyzed meteorites. CK chondrites exhibit similar spectra to CM, but higher overall reflectance probably due to less pervasive aqueous alteration. CKs are in general darker than pure olivine and exhibit its 1.05 µm absorption band, also having blue sloped spectra (Cloutis et al., 2012e). Figure 1b shows the easily distinguishable red-sloped spectra of the CR and R groups.

In general chondrite groups unaffected by aqueous alteration (e.g. CO and CV) are dominated by the features of silicates. They are the most abundant minerals forming the components of chondrites, so they exhibit diagnostic features distinguishable in the visible and near-infrared region. For example, crystal field absorptions appear in the spectra due to the presence of transition metal ions (often $Fe^{2+}$) in the mafic minerals like e.g. olivine and pyroxene. In this sense, olivine is characterized by a typical absorption feature at 1 µm, while pyroxene shows two absorption bands at 1 and 2 µm. On the other hand, the 1.2 µm band is obvious in Fig. 1c, and corresponds to olivine (see Table 3). Fig. 1c also shows that most of the analyzed meteorites except Allende (that it is a fall) exhibit decreasing slopes below 500 nm that are probably consequence of iron oxyhydroxides produced by terrestrial alteration. In fact, Buchwald and Clarke (1989) identified the primary Fe-bearing terrestrial weathering products in Antarctic chondrites are akaganéite (β-FeOOH), goethite (α-FeOOH), lepidocricite (γ-FeOOH), and maghemite (γ-$Fe_2O_3$). It is particularly remarkable the narrow absorption seen in ALHA77003 and Kainsaz that we think could be associated with maghemite and magnetite due to its deep extinction below 400 nm (see e.g. Tang et al., 2003). In fact, this deep absorption feature around 300 nm is also visible in Fig 1a for CK4 ALH85002 and CK4/5 PCA82500 as well as in the altered CM2 chondrites: MAC 02606 and SCO 06043. Also the spectral profiles of CR2 and R3 in Fig. 1b are indicative of the abundant presence of iron oxyhydroxides as previously shown (Cloutis et al., 2012a).

In general, an asteroid reflectance spectrum affected by aqueous alteration may be distinguished by several features (Dotto et al. 2005). Phyllosilicates show absorption bands centered around 0.7 µm, attributed to $Fe^{2+}\rightarrow Fe^{3+}$ charge transfer in iron oxides (Cloutis et al. 2011a,b). Pervasive water alteration changes the bulk mineralogy transforming Fe-Ni grains into magnetite and troilite. Magnetite for example has distinctive absorption bands in 0.48 and 1 µm respectively (Cloutis & Gaffey, 1994; Cloutis et al. 2011b; Yang & Jewitt, 2010). Fornasier et al. (1999) also found a relationship between the albedo of the objects and the extent of aqueous alteration that can be explained with the progressive leaching of iron from silicates as the alteration proceeds. In fact, Vilas (1994) remarked that leached iron, the most opaque phase in the visible range associated with aqueous alteration, is enveloped in magnetite and iron sulfide grains, so much less material would be available to absorb the incoming sunlight and this would naturally increase the albedo. Yang and Jewitt (2010) also identified the magnetite absorption band as a distinctive feature in B-type asteroids.

The highest reflectance in our data sample is achieved by CCs belonging to the CR, R and CH groups. Figure 1b shows the characteristic red-sloped, and widely varying spectra of CR and R groups ranging from about 5 to 20% maximum reflectance. The three studied CRs show distinctive magnetite absorption bands in 0.48 and 1 µm. These groups don't exhibit the absorption band in the 650 nm region characteristic of CMs attributable to $Fe^{3+}$-$Fe^{2+}$ charge transfers, as also noted by Cloutis et al. (2012a). This suggests that CMs are richer in $Fe^{3+}$ phyllosilicates and aqueous alteration was probably weaker in most CRs. Fig. 1b also shows weak absorption bands in the 900-1,300 nm region, as also was found by Cloutis et al. (2012a). We initially interpret the largest reflectance of CR, R, and CHs compared to other chondrite groups as probably due to their major abundances of metal grains in volume, but alternative explanations exist. For example, Gaffey (1986) showed that cut metal grains are redder than just crushed metal grains, so the red slope is probably due to the fact that the metal grains in our studied sections are cut, therefore the red slope argument is probably not valid. On the other hand, Cloutis et al. (2012a) found that due to having low iron content in their silicates, even small amounts of terrestrial weathering products will cause a red slope. In other words, this non-linear relationship between metal content and reflectivity needs to be better explored, having also into account the different availability of metal. For example, CR chondrites have most of the metal inside chondrules while CH ones have them in the matrix. We expect that in nature chondrules are not perfectly cut by space weathering processes (unlike happens in thin or thick sections) and consequently metal grains inside of chondrules will not contribute to the spectrum as much as they do in the matrix.

It is also remarkable that CR, CH and CB chondrites are considered primitive meteorites forming a grouplet sharing similar characteristics and being particularly rich in metal (Weisberg et al. 2006). The metal content was estimated to be respectively 5-8, 20, and 60-80 in vol.% for the above mentioned chondrite groups (Scott and Krot 2005; Weisberg et al. 2006). We have found recently that aqueous alteration in CRs decreases the abundance of metal grains producing magnetite aggregates and sulphides (Trigo-Rodríguez et al. 2013). Consequently, as noted by Moyano-Cambero et al. (2013), aqueous alteration may be a natural pathway to decrease the reflectivity of the parent bodies of CR and CH chondrites as exemplified in Fig. 6. We have not included in the discussion here the CBs as they are probably formed by impacts, but we include in Fig. 1d the reflectivity of CBb chondrite MAC 02675, ranging from 30% in the UV to about 47% at 2 µm.

CO and CV groups also can have high reflectance, and some COs like e.g. ALH 83108 and ALHA 77003 reflect more than the 10% of the light at wavelengths longward of 500 nm. However these Antarctic finds are showing pervasive, spectrally significant, terrestrial weathering due to the oxidation of metal and mafic silicates (Bland et al., 2006). This produces in both meteorites an increase in the UV region absorption, and the visible region absorption edge appears shifted to longer wavelengths (Cloutis et al., 2012e), as well as the appearance of an absorption feature near 0.9 µm due to $Fe^{3+}$ charge transfers (Sherman, 1985). Specimens from the CO and CV groups are plotted in Fig. 1c, and mostly characterized by an overall flat spectrum until about 2,000 nm and a reddish growing slope onwards. Both groups spectra also shows weak absorption bands in the 900-1,300 nm region as was found by Cloutis et al. (2012a,c). It has been described that the COs have a ubiquitous absorption feature in the 1 µm region, and other in the 2 µm for petrologic subgrades >3.1 that have experienced increasing degrees of thermal metamorphism (Cloutis et al., 2012c).

Finally, Fig. 1d compiles all reflectance spectra obtained so far where we have also included CH and CB chondrites. At the given Y-axis scale the low-reflectance groups discussed in previous figures appear overlapped, well exemplifying their dark nature and constrained albedos. We clearly note that CH and CB chondrites (represented respectively here by PCA 91467 and MAC 02675) seem to be far more reflective than the rest of CCs that are in the range of 5-15% reflectance in the full wavelength window. This is consequence of the increasing abundance of metal in CH and CB chondrites. In fact, CR, CH and CB chondrites are primitive meteorites forming a grouplet that share similar characteristics and particularly rich in metal (Weisberg et al., 2006). These groups probably recorded high-temperature condensation processes in the early protoplanetary disk (Weisberg et al., 2001, 2006). In any case, as we mentioned before, some members of the CH group exhibits clear features of extensive aqueous alteration. We have presented here just the spectrum of only one CH chondrite (PCA 91467) using a thin section that exhibits moderate terrestrial alteration over the very-near meteorite surface, but seems quite unaltered in depth. We just put the beam over the relatively pristine region that seen to the microscope is not showing significant amounts of terrestrial alteration oxides. It then probably difficult to compare with the spectrum presented by Cloutis et al. (2012f) in their Fig. 10a as we don't know the origin of the powders, but we notice that are quite similar except that the reflectance seems to decrease more for powders in the UV window. As they, we think that additional work on CHs exhibiting different degrees of terrestrial weathering is required to reach significant conclusions.

4.2. IR spectra: clues on clay minerals and organics.

IR spectra obtained by the ATR technique are relevant to provide a general assessment of the main mineral bands and organic features contained in grinded powders of chondrites (Trigo-Rodríguez et al., 2012). The most relevant features observed in our spectral data are briefly summarized in Table 4, but a list of bands and features can be found in Pretsch et al. (2002, 2009). Overall graphs showing the spectra of different chondrite groups are presented in Figures 2-5. Spectral similarity among members of a same CC group is found, but still some obvious differences exist. The most significant difference for specimens belonging to a same chondrite group is the different depth in the OH stretching hydroxyl bands, the Al/Si-OH libration bands, and other functional bands (see Table 4 for relative band depths). Such differences appear correlated, and suggest different extent of aqueous alteration and adsorbed water in their forming minerals as we previously found for CMs (Rubin et al., 2007). On the other hand, to compare the spectra we have separated the data in three different figures (Fig. 2-5). For clarity the chondrite groups plotted in such figures are compiled according to groups exhibiting a similar location in their Reststhralen features and absorption peaks.

4.3.1. IR spectra of CI and CM chondrites.

CI and CM groups are with difference the most aqueous altered chondrites (see e.g. Zolensky & McSween, 1988; Brearley & Jones, 1998). Figure 2 shows a compilation of CM chondrites compared with the Orgueil CI chondrite. Orgueil could be considered a good proxy as it represents a CI that is more hydrated than any CM, and consequently exhibits much deeper absorption bands.

The key 3 μm band region is usually used to quantify the presence of bonded water in phyllosilicates ((Lebofsky 1978, 1980; Vilas & Gaffey 1989). The OH absorption band is considered a quite distinctive and useful feature to distinguish aqueously altered asteroids (Rivkin and Emery, 2010). Recent work using ATR spectroscopy has demonstrated the presence of several bands for $H_2O$ at about 16.7, 6.2, 5.9, and 2.8 μm (Maréchal, 2011). These bands are observed in our ATR spectra of aqueously altered chondrites (Tables 4 and 5). It is important to remark that in liquid water the molecular stretch vibrations shift to higher frequency on raising the temperature. The reason is as H-bond weakens, the covalent O-H bonds strengthen as consequence of high-frequency induced vibration (Praprotnik et al., 2004). In general, the O-H bond stretch frequency at surfaces where the water molecule has 3 H bonds occurs at ~2.7 μm. Consequently, the positions and width of the absorption bands are temperature-dependent, but not important differences are seen in our spectra as the temperature was kept stable in our laboratory to ~20 ºC. For example, a OH absorption band was found by Rivkin et al. (2012) in Marco Polo-R backup target candidate NEO 1996 FG3, concluding its plausible association with the CM chondrites.

Our IR spectra of CM chondrites exhibit very different aqueous alteration degrees as is expected from mineralogical and petrological clues (Rubin et al., 2007). Some specimens like e.g. QUE 99355, LEW 87148, and Murray exhibit quite deep OH band, but Cold Bokkeveld, QUE 97990, and MET 01070 exhibit a moderate OH band. while finally MAC 02606 has a very weak band. In all cases samples were crushed in an agate mortar and the resulting powder was immediately placed in the IR spectrometer without any further manipulation to minimize moisture absorption. However, our measured OH bands (see Fig. 2b) are wider than those obtained after heating and measuring under vacuum conditions by other authors, indicating some water absorption, although the maximum IR absorption peaks are not significantly displaced (see e.g. Takir and Emery; Takir et al. 2012). In any case, the rounded OH absorption bands that we obtain (see magnified Fig. 2b, online) are also similar to the obtained reflectance spectra of asteroids like 361 Bononia (see Fig. 6 of Takir and Emery 2012). They presented evidence that some primitive asteroids have distinctive sharp and rounded OH absorption bands that could be used for asteroid sub-classification.

It is remarkable that MAC 02606 spectrum exhibits a weak OH absorption band like also occurs for the aqueously altered CM1 chondrites analysed: MIL 07689 and SCO 06043. These two CM1 chondrites have not noticeable OH absorption band at 3.1 μm. This quite surprising result, could suggests that extremely altered CM chondrites can have been naturally heated losing part of the bonded water. Looking for more details in the Meteoritical Bulletin on MIL 07689 and SCO 06043 we found that they are small meteorites (12.2 and 27.6 g in mass respectively) that were quite fractured during atmospheric entry. We suspect that it could be produced by shock compression during atmospheric entry as consequence of the overload pressure suffered during their deceleration. As the shock propagates through the target and part of the energy is expanded to collapse the empty spaces some compression occurs (Britt et al. 2002). In fact, we know that surviving chondrites are probably biased towards the high strength portions of the incoming meteoroids, and compression associated with atmospheric penetration usually leads to meteoroid disruption in the atmosphere (Trigo-Rodríguez and Blum 2008). Also supporting the quality of our IR data Fig. 2b shows that the CM chondrite falls (Cold Bokkeveld, Murchison, and Murray) exhibit quite different OH band depths independently of their terrestrial residence time. This again suggests that (terrestrial) absorbed water is probably playing a minor role here. The degree of aqueous alteration in CM chondrites, and its impact in the modal mineralogy determined by X-ray diffraction has been recently studied (Howard et al., 2009, 2011).

The spectral region where silicate stretching occurs (~10 μm) and the associated with bending modes occurring at wavelengths between 15 and 25 μm are particularly diverse for the selected CMs. This is quite important as we remark that could be used to discriminate the presence of pristine mafic silicates from the most common phyllosilicates available in more aqueously altered chondrites. We remark that the olivine exhibits reflectance minima at 11.2 and 19.5 μm, while the phyllosilicates show a reflectance minimum due to $SiO_4$ stretching at about 10 μm, having bands at about 16 and 22 μm due to the bending vibration of the hydroxyl group. CM chondrites moderately aqueous altered exhibit reflectance minima associated with mafic silicates (see Fig. 2a) as well exemplified by QUE 97990, Murchison, Murray or QUE 99355 IR spectra. We remark that the presence of an olivine feature at about 11.2 μm for petrologic subtypes larger than 2.3 is consistent with Rubin et al. (2007) aqueous alteration sequence. The most extensively altered CM chondrites are not showing such band.

Another band feature of CM spectra is the different depth found for the ~10.5 μm Al/Si-OH libration absorption band that is probably characteristic of the amount of phyllosilicates present. In fact, when we compare QUE 99355 and Murray from one side, and QUE 97990 and LEW87148 it can be noticed that when the olivine peak disappears the relative depth of the 10.5 μm band increases. Even the absorption band exhibited by QUE 99355 is comparable with that one of the extensively aqueous altered CI chondrite Orgueil where mafic silicates are gone due to be extensively altered (Brearley and Jones, 1998). A very broad band around 16.4 μm due to the Al-O and Si-O out of plane bonds is also shown for CI and CM spectra. In Orgueil, such band extends until 17.6μm due to the possible action of other minerals.

CI and CM chondrites are also very rich in organic-related absorption features. Some are narrow features, but other are wide bands caused by main-forming minerals like e.g. clays. In fact, many of the bands tentatively identified in Table 4 and 5 are associated with organic bonds. For example, the 6.1μm CC double bond stretch is very strong in CI chondrite Orgueil, being more moderate but still discernible in CMs (see Fig. 2a,b).

Finally, when plotting CM chondrites we found one specimen called QUE 99038 that exhibits differen**ces** from the characteristic features of CO and CV chondrites. The Meteoritical Bulletin currently provides a weird description of that chondrite, probably suggesting that it has anomalous composition: "This carbonaceous chondrite has a dark gray to blackish crust, but appears to be a melted glob of chondrules spread over the entire exterior of the rock. The surface has a vesicular-like texture. The interior is charcoal grey with numerous mm-sized chondrules and light and dark grey crystalline materials. This meteorite is moderately hard and has a sulphurous odour". All this evidence suggests that QUE 99038 could be misclassified (Rubin, pers. comm.), so we have removed it from Figure 4.

4.3.2. IR spectra of CO and CV chondrites.

The spectra of CO and CV chondrites are included in Fig. 3. These carbonaceous chondrites have strong olivine absorption bands compared with those of CIs and CMs because mafic silicates have been preserved in CO and CV groups. In fact, the main absorption bands of CO and CV are displaced to larger wavelengths as a consequence of significant differences in mineralogy. CO and CV chondrite groups have been subjected to little or no aqueous alteration so their minerals are mostly unaltered. The deep olivine 11.6 μm absorption band exhibits other two narrow bands at 9.4 and 10.5 that are respectively distinctive of the N-C and the Al/Si-OH libration bonds. Refractory oxides as e.g. hibonite are minerals abundant in Ca-Al-rich inclusions and presolar grains (Brearley & Jones, 1998) that probably produce the feature at 12.1 μm, with probably a minor contribution from the C-H bend inherent to organic compounds. The 17.3μm band is also characteristic of CO and CV chondrites, being narrower and displaced to longer wavelength than the 16.4 μm band typical of CMs.

CO and CV chondrites also show a broad band around 26.0 μm that corresponds to a crystalline silicate: fosterite. Such feature is not seen in the aqueous altered chondrite groups as we pointed out previously. This is probably due to the pervasive effect of aqueous alteration in crystalline silicates that are transformed into clay minerals for the most altered chondrite groups.

4.3.3. IR spectra of other chondrite groups: R, CK, CR, and CB.

Finally, in Fig. 4 we compiled the IR spectra of rare groups of carbonaceous chondrites: CB, CH, CK and R specimens. Deep absorption peaks for CK and R chondrites are located at 10.4, 11.6 and 12.1 µm, and are also common to CO and CVs (see Fig. 2). These are associated respectively with the Al/Si-OH libration band of phyllosilicates, the olivine reststrahlen band, and to refractory oxides (see Table 5). On the other hand, the reflectance spectra of the R3 and the CK4 chondrites analysed are remarkably similar to the exhibited by CV3 Allende.

5. Conclusions.

UV-Vis to far IR reflectance spectra of CCs have been presented using two different spectrometers. Detailed reflectance spectra were obtained for specimens belonging to the CM, CO, CH, CB, CK, and R groups. Most groups exhibit distinctive slopes and reflectivity features in the UV to NIR window, while they show characteristic absorption bands of silicates and phyllosilicates dominating the spectra in the IR window. All these reflectance patterns are useful to remotely characterize the surface mineralogy of primitive carbon-rich asteroids. To define future targets to be explored by spacecraft the CM, CO, CH and CK chondrites exhibit OH absorption bands characteristic of phyllosilicates. The different relative depths are probably consequence of different degrees of aqueous alteration in agreement with previous studies using IR micro-spectroscopy (Beck et al., 2010) or using mineral patterns found in CMs (Rubin et al., 2007).

A careful identification of absorption features and the location of other absorption bands (e.g. these located around 10, 17, 20 and 26 µm) could be used to differentiate between carbonaceous asteroids and establish a more precise relationship between asteroid taxonomy and primitive meteorites arrived to Earth. In case that future in situ exploration of primitive asteroids will be achieved, this study can contribute to establish additional criteria to distinguish more pristine regions to be sampled back to Earth. Such is the main goal of several space missions like Osiris Rex, Hayabusa 2, and, if selected for flight, Marco Polo-R.

The main conclusions of the present work are:

i) Laboratory spectra of carbonaceous chondrites are extremely useful to identify the main patterns and absorption features that can be used to remotely characterize primitive asteroids. A complete spectral characterization from 0.2 to 30 µm provides distinctive features that can be used for such purpose.

ii) The OH absorption band characteristic of water bounded in phyllosilicates is usually present in CM, CO, CH and CK chondrites, but its relative depth is highly variable and probably consequence of different degrees of parent body aqueous alteration, though it may also be affected by terrestrial weathering as is evidenced by some of our spectra of Antarctic finds.

iii) In general polished slabs spectra are useful to identify the main absorption band wavelength positions, but are not fully representative of the asteroids reflectance spectra due to the presence of powdered regoliths in most asteroidal surfaces.

iv) A careful identification of IR absorption features and their location and relative depth seems particularly useful to characterize asteroids. Particularly the major or minor presence of water bands and bonds associated with organics could be used to characterize targets selected for future spacecraft missions.

v) High-resolution ATR IR spectra of CCs are useful to provide a general assessment of the main mineral bands and organic features contained in the ground powders. The identified bands and features can be used to identify similar patterns in dark primitive asteroids at long IR wavelengths.

vi) The non-destructive character of the ATR measurements, and the really minimum mass of meteorite powders required to get an accurate IR spectrum (~0.02 g) makes this a promising technique to characterize future sample returned materials. Particularly, the OH band conclusions could be more significant if future studies can be made under vacuum.

vii) The IR spectra are also useful to find chemical and mineralogical differences among the specimens belonging to the same chondrite group. The ATR technique has been proved to be useful to discover misclassified chondrites or having anomalous composition.

Acknowledgements

Current research was supported by the Spanish Ministry of Science and Innovation (project: AYA2011-26522) and CSIC (starting grant #201050I043). J.Ll. is grateful to ICREA Academia program. Z.M. acknowledges support from the Royal Society. NASA Meteorite Working Group and Johnson Space Center meteorite curators are acknowledged for providing the Antarctic carbonaceous chondrites. We also want to express our sincere gratitude for the valuable effort made over the years in the recovery of Antartic samples by ANSMET (The Antarctic Search for Meteorites program). Katsuhito Othsuka and R.H. are also acknowledged for providing some of the carbonaceous chondrites from historic falls.

. REFERENCES

# TABLES

| Asteroid taxonomic class | Example of asteroids belonging to the class | Suggested chondrite group | Possible meteorite in our sample | Spectral properties: (slope; albedo) | Main spectral features |
|---|---|---|---|---|---|
| A | 446 Aeternitas | R3 | PRE 95404 | Red spectrum below 0.4 µm | Absorption by olivine |
| B | 2 Pallas | CM/CR | SCO 06043 | Blue in Vis; albedo usually > 0.1 | Weak UV feature |
| C | 10 Hygiea, 324 Bamberga | CM/CI | MAC 02606 | Flat to reddish below 0.4 µm; albedo < 0.1 | Weak UV feature |
| D | 152 Atala, 624 Hektor, 773 Irmintraud, 944 Hidalgo | CI | QUE 99355 | Reddish spectra | Featureless and with very low albedo |
| F | 704 Interamnia | CR – CM | MET 01079 | Flat to bluish below 0.4 µm; albedo < 0.1 | VIS wavelength only as DeMeo et al. 2009 found it diverging in IR. Very weak UV feature |
| G | 1 Ceres | CM1 | Murchison | Flat past 0.4 µm; low albedo | Defined only for UV-visible region as its distinguishing feature is in the UV. Strong UV feature; usually strong 3 µm OH band |
| K | 181 Eucharis, 221 Eos, 402 Chloë | CO3/CV3 | MET 01074 | Shallow 1 µm feature and lack of 2 µm absorption | Intermediate between C and S asteroids |
| L | 83 Beatrix | CV3 | Allende | Flat longwards 0.75µm, but strongly reddish below it | Very strong UV |
| P | 46 Hestia, 65 Cybele, 76 Freia, 87 Sylvia, 153 Hilda | CI/CM | Orgueil | Flat to slightly red; albedo <0.05 | Very low albedo and featureless spectra |

Table 1. Asteroid classes and temptative meteorite analogs as deduced from the observed UV to IR features. Only primitive taxonomic classes are included in this table. Reflectance classes are taken from Tholen (1989) and DeMeo et al. (2009). Some features from UV to NIR are described from Burbine (2002), and de León et al. (2012).

| Meteorite name | Chondrite group/petrologic subtype | Weathering grade | Year of find (F, for year of fall) | Spectra obtained in this study |
|---|---|---|---|---|
| ALHA77003 | CO 3.6 | Ae | 1977 | UV-NIR |
| ALH 83108 | CO 3.5 | A | 1983 | UV-NIR + IR |
| ALH 84028 | CV 3 | Ae | 1984 | UV-NIR |
| ALH 85002 | CK 4 | A | 1985 | IR |
| Allende | CV 3 | A | F 1969 | UV-NIR + IR |
| Cold Bokkeveld | CM 2 | A | F 1838 | IR |
| EET 92159 | CR 2 | B/C | 1992 | UV-NIR |
| GRA 95229 | CR 2 | A | 1995 | UV-NIR |
| Kainsaz | CO 3.2 | A | F 1937 | UV-NIR + IR |
| LAP 02342 | CR 2 | A/B | 2002 | UV-NIR |
| Leoville | CV 3 | A/B | 1961 | IR |
| LEW 87148 | CM 2 | Ae | 1987 | UV-NIR + IR |
| MAC 02606 | CM 2 | A | 2002 | UV-NIR + IR |
| MAC 02675 | CBb | B | 2002 | IR |
| MET 01070 | CM 1 | Be | 2001 | UV-NIR + IR |
| MET 01074 | CV 3 | B | 2001 | UV-NIR |
| MIL 07689 | CM 1 | C | 2007 | UV-NIR + IR |
| Murchison | CM 2 | A | F 1969 | UV-NIR + IR |
| Murray | CM 2 | A | F 1950 | IR |
| Orgueil | CI 1 | B | F 1864 | IR |
| PCA 91467 | CH 3 | B/C | 1991 | UV-NIR + IR |
| PRE 95404 | R3 | A | 1995 | UV-NIR + IR |
| QUE 97990 | CM 2 | Be | 1997 | UV-NIR + IR |
| QUE 99038 | CM 2 | A/B | 1999 | UV-NIR + IR |
| QUE 99355 | CM 2 | B | 1999 | UV-NIR +IR |
| SCO 06043 | CM 1 | ~B | 2006 | UV-NIR + IR |
| Tagish Lake | C 2 Ungrouped | A | F 2000 | UV-NIR |

Table 2. Selection of carbonaceous chondrites analyzed in this work. The main grain size measured for the grained meteorite powders is 100±50 μm.

| Meteorite name | Chondrite group/petrologic subtype | Olivine Fa content range (average) | Pyroxene Fs content Range (average) | Notes and secondaryminerals | Reference |
|---|---|---|---|---|---|
| ALHA77003 | CO 3.6 | 4-48 (22) | 2-25 (14) | Pyroxene contains ~1% CaO, spinel grains | Met. Bull. #76 |
| ALH 83108 | CO 3.5 | 0.9-38 | 1-17 | Pyroxene compositions range from pure MgSiO3 to Wo5Fs16 | Met. Bull. #76 |
| ALH 84028 | CV 3 | 0-50 | 2 | Large CAIs with gehlenitic melilite, spinel and Ti-rich fassaitic pyroxene | Met. Bull. #76 |
| ALH 85002 | CK 4 | 30 | 26 | Plagioclase: An54-59 | Met. Bull. #76 |
| Allende | CV 3 | 47 | 1-4 | Fo: 53. Oxidized group exhibiting: magnetite, sulfides, metal, nepheline, sodalite,... | Peck (1983); Krot et al. (1998); Brearley & Jones (1998) |
| Cold Bokkeveld | CM 2 | n.a. | n.a | Mg-rich serpentine + olivine | Howard et al. (2008) |
| EET 92159 | CR 2 | - | - | Metal and pyrrhotite replacement by magnetite | Trigo-Rodríguez et al. (2013) |
| GRA 95229 | CR 2 | 1-31 | 2-4 | Low vol% of matrix | Met. Bull. #82 |
| Kainsaz | CO 3.2 | - | - | Plagioclase-rich chondrules | Brearley & Jones (1998) |
| LAP 02342 | CR 2 | 0-5 | 1-3 | - | Met. Bull. #89 |
| Leoville | CV 3 | 0-7 (1) | n.a. | CAIs and AOAs | Komatsu et al. (2001)- |
| LEW 87148 | CM 2 | 0-22 | 2-58 | No troilite or Fe-Ni | Met. Bull. #76 |
| MAC 02606 | CM 2 | 0-2 | - | Rare sulfide grains | Met. Bull. #88 |
| MAC 02675 | CBb | - | 1-4 | 70-80% metal particles | Met. Bull. #89 |
| MET 01070 | CM 1 | Be | 2001 | Highly altered: Fe-rich serpentine, rare sulfide grains | Met. Bull. #87 |
| MET 01074 | CV 3 | 0-9 | 0-2 | Chondrules and CAIs set in dark matrix | Met. Bull. #88 |
| MIL 07689 | CM 1 | n.a. | n.a. | Fe-rich serpentine | Met. Bull. #99 |
| Murchison | CM 2 | 1-60 (5) | n.a. | phyllosilicates | Hutchison & Symes (1972); |
| | | | | | Brearley & Jones (1998) |
| Murray | CM 2 | 1-50 (5) | - | phyllosilicates | Hutchison & Symes (1972); Brearley & Jones (1998) |
| Orgueil | CI 1 | ~2 | n.a. | phyllosilicates | Brearley & Jones (1998) |
| PCA 91467 | CH 3 | 1-58 (3) | 1-46 (3) | Metal-rich | Met. Bull. #76; Bischoff et al. (1994) |
| PRE 95404 | R3 | 1-41 | 7-21 | - | Met. Bull. #82 |
| QUE 97990 | CM 2 | 0-61 | 0-1 | Spinel, pyroxene, olivine | Met. Bull. #85, Rubin (2007) |
| QUE 99038 | CV 3 | 1-39 | - | Pentlandite, troilite, magnetite | Met. Bull. #85 |
| QUE 99355 | CM 2 | 1-42 | - | Fe-rich serpentine in matrix plus some metal and sulfide | Met. Bull. #86 |
| SCO 06043 | CM 1 | n.a. | n.a. | - | - |
| Tagish Lake | C 2 Ungrouped | 1 | n.a. | Olivine (Fo$_{99}$), magnetite and phyllosilicates | Zolensky et al. (2002) |

Table 3 (**online**). Main mineralogical properties of the analyzed meteorites. Abbreviations: Fa:fayalite, Fo:forsterite, Fs:ferrosilite, CAIs:Ca-Al rich inclusions, AOAs: Ameboid Olivine Aggregates, and n.a.: non available.

| Meteorite name | Chondrite group | Maximum reflectance | Main identified absorption bands λ(μm) | | | | | | | | Reflectance in band minimum | | | | | | | Band assignment |
|---|---|---|---|---|---|---|---|---|---|---|---|---|---|---|---|---|---|---|
| ALH 77003 | CO 3.6 | 91.3 (at 8.3μm) | 3.8 | 10.5 | 11.6 | 12.1 | 17.3 | 21.2 | | | 88.6 | 65.3 | 45.9 | 51.9 | 64.4 | 40.3 | | AS/CA/CR/HI/CO/HYD |
| ALH 83108 | CO 3.5 | 92.6 (at 8.3μm) | 3.1 | 6.1 | 9.4 | 10.5 | 11.6 | 12.1 | 17.1 | 21.1 | 89.2 | 90.6 | 86.5 | 65.7 | 46 | 50.7 | 64.6 41 | OH/HCC/NC/AS/CR/HI/CO/H |
| ALH 85002 | CK 4 | 90.1 (at 8.1μm) | 4.31 | 10.4 | 11.5 | 12.1 | 17.1 | 20.9 | | | 68.7 | 55.6 | 58.5 | 67.1 | 59.7 | 46.1 | | CO$_2$/AS/CR/HI/CO/HYD |
| Allende | CV 3 | 87.2 (at 8.8μm) | 4.3 | 9.4 | 10.4 | 11.6 | 12.1 | 17.2 | 21.2 | | 83.5 | 80.4 | 53.4 | 28.5 | 31.7 | 53.4 | 25.3 | CO$_2$/NC/AS/CR/HI/CO/HYD |
| Cold Bokkeveld | CM 2 | 93.3 (at 8.1μm) | 3.1/ | 4.3 | 6.2 | 7.1 | 7.6 | 10.4 | 16.6 | | 91.8 | 93.9 | 91.5 | 92.6 | 93.3 | 63.9 | 66.7 | OH/CO$_2$/HCC/CH$_2$/AS/HOX |
| Kainsaz | CO 3.2 | 89.9 (at 2.8μm) | 3.1 | 3.8 | 4.3 | 9.4 | 10.5 | 11.5 | 17.1 | 21.0 | 88.6 | 88.6 | 86.6 | 77.3 | 65.9 | 53.7 | 67.4 47.7 | OH/CA/CO$_2$/NC/AS/CR/HI/CO/HYD |
| Leoville | CV 3 | 89.8 (at 8.3μm) | 6.8 | 9.3 | 10.5 | 11.5 | 12.0 | 20.9 | | | 88.8 | 81.5 | 74.1 | 68.3 | 70.9 | 57.3 | | NC/AS/CR/HI/CH$_4$ |
| LEW 87148 | CM 2 | 87.4 (at 8.1μm) | 3.1 | 4.3 | 6.2 | 10.6 | 16.5 | | | | 80.4 | 84.2 | 83.4 | 43.8 | 51.1 | | | OH/CO$_2$/HCC/AS/HOX |
| MAC 02606 | CM 2 | 94.7 (at 8.3μm) | 3.1 | 4.5 | 6.2 | 7.1 | 10.1 | 11.5 | 16.6 | | 94.8 | 93 | 93.7 | 90.4 | 78.3 | 80 | 78.6 | OH/HCC/CH$_2$/AS/CR/HOX |
| MAC 02675 | CBb | 97.9 (at 8.3μm) | 9.4 | 10.6 | 11.6 | 12.0 | 19.9 | 24.0 | | | 95.4 | 93.8 | 93.1 | 93.5 | 87 | 84.1 | | NC/AS/CR/HI/OL/PAH |
| MET 01070 | CM 1 | 98.6 (at 8.4μm) | 3.0 | 4.5 | 6.2 | 7.1 | 10.6 | 16.3 | 24.0 | | 88.7 | 89.1 | 89.1 | 66.4 | 67.1 | 47.8 | | OH/HCC/AS/HOX/PAH |
| MIL 07689 | CM 1 | 98.6 (at 8.3μm) | 3.2 | 4.3 | 10.5 | 15.8 | | | | | 98.2 | 98.0 | 95.4 | 94.1 | | | | OH/CO$_2$/AS/? |
| Murchison | CM 2 | 91.7 (at 8.3μm) | 3.0 | 4.5 | 6.2 | 7.1 | 10.7 | 16.3 | 24.0 | | 88.7 | 89.1 | 89 | 89.5 | 66.6 | 67 | 47.1 | OH/SiO/HCC/CH$_2$/AS/HOX/PA |
| Murray | CM 2 | 81.1 (at 8.1μm) | 3.1 | 4.4 | 6.2 | 7.1 | 10.6 | 11.6 | 16.4 | | 76.1 | 80.1 | 77.3 | 78.5 | 37 | 42.2 | 47.5 | OH/SiO/HCC/CH$_2$/AS/CR/HOX |
| Orgueil | CI 1 | 80.9 (at 8.1μm) | 3.2 | 6.2 | 9.3 | 10.5 | 16.5 | 17.1 | 23.9 | | 61.4 | 69.6 | 50.6 | 34.4 | 23.9 | 14.1 | 13.9 | OH/HCC/NC/HOX/AS/PAH |
| PCA 91467 | CH 3 | 91.2 (at 8.1μm) | | 11.7 | | 20.3 | | | | | | 72.8 | | 60.2 | | | | CR/CH$_4$ |
| PRE 95404 | R3 | 90.6 (at 8.1μm) | 3.1 | 6.1 | 9.3 | 10.4 | | | | | 97.9 | 88.8 | 80.7 | 62.7 | | | | OH/HCC/HOX/AS |
| QUE 97990 | CM 2 | 89.6 (at 8.3μm) | 3.1 | 4.3 | 6.2 | 10.7 | 11.5 | 16.6 | 20.0 | | 80.2 | 83 | 83.1 | 50 | 52.3 | 47.6 | 40.3 | OH/SiO/HCC/AS/CR/HOX/OL |
| QUE 99038 | CM 2 | 91.1 (at 8.2μm) | 3.0 | 4.5 | 6.0 | 10.4 | 11.5 | 12 | 17 | 19.7 | 85.2 | 85 | 87.8 | 57.8 | 34.5 | 37.1 | 57 38.8 | OH/SiO/HCC/AS/CR/HI/CO/HYD/FO |
| QUE 99355 | CM 2 | 86.4 (at 8.1μm) | 30.0 | 6.1 | 10.5 | 16.5 | 20.1 | 24.1 | | | 74.3 | 80.6 | 33.2 | 42.1 | 37.1 | 16.6 | | OH/HCC/AS/HOX/OL/PAH |
| SCO 06043 | CM 1 | 95.9 (at 8.3μm) | 3.1 | 6.2 | 9.1 | 10.5 | 16.2 | 23.7 | | | 95.2 | 95 | 93.1 | 86.4 | 85.4 | 74.4 | | OH/HCC/NC/AS/HOX/PAH |

Table 4 (**online**). Main absorption bands in IR found for the analyzed chondrites. Relative band depth can be estimated from the maximum reflectance. Band abbreviations are listed in Table **5**.

| Band or feature | Abreviation in Table 4 | λ (cm$^{-1}$) | λ (μm) | Appearance in (Location: chondrite groups) |
| --- | --- | --- | --- | --- |
| OH stretching hydroxyl groups | OH | 3,450 | 2.9 | Phyllosilicates: CM, CI, CR, R |
| CH stretch band | CH | 3,226-3,030 | 3.1-3.3 | Organics in matrix: CI, CM, R |
| Carbonates | CA | 2,632 | 3.8 | Carbonates + minor contribution of $C_2H_2$ |
| $CO_2$ overtone | $CO_2$ | 2,353 | 4.25 | $CO_2$ vibrating group (from the atmosphere) |
| Si-O bond | SiO | 2,247 | 4.45 | Si-O rich glass: CM |
| $H_2O$ / CC double bond stretch | HCC | 1,650 | 6.1 | Water + organics: CI, CM, R |
| $CH_2$ & $CH_3$ bend bands | $CH_2$ | 1,450 & 1,400 | 6.9 & 7.1 | Organics in matrix: CM |
| N-C stretch | NC | 1,066 | ~9.4 | Organics in matrix: CO and CV |
| Al/Si-OH libration bands | AS | 930-950 | 10.8-10.5 | Variable location |
| Reststhralen features and Christiansen peak | CR | ~892-865 | ~11.2-11.6 | Olivine feature: variable as consequence of differences in grain size, and mineralogy |
| Hibonite | HI | ~825 | ~12.1 | Hibonite: characteristic of pristine mineralogy in CO, CVs. Also a minor contribution of the C-H bend in aromatics |
| $H_2O$ absorption band and Al-O and Si-O, out of plane | HOX | ~600 | ~16.7 | Phyllosilicates: water absorption band characteristic of: CI, CM |
| $CO_2$ absorption band | CO | ~577 | ~17.3 | Organics in matrix: CM, CO, CV |
| Olivine absorption peak | OL | ~502.5 | ~19.9 | Olivine |
| Hydroxyl group | HYD | ~472 | ~21.2 | Bending vibration of hydroxyl group: CI, CM, CV |
| Fosterite ($Mg_2SiO_4$) | FO | ~383 | ~26.1 | Silicate-rich groups: CO, CV |

Table 5. A selection of the main absorption bands and sharp features identified in ATR IR spectra of CCs. Pretsch et al. (2002, 2009), Mustard & Hays (1997).

# FIGURES

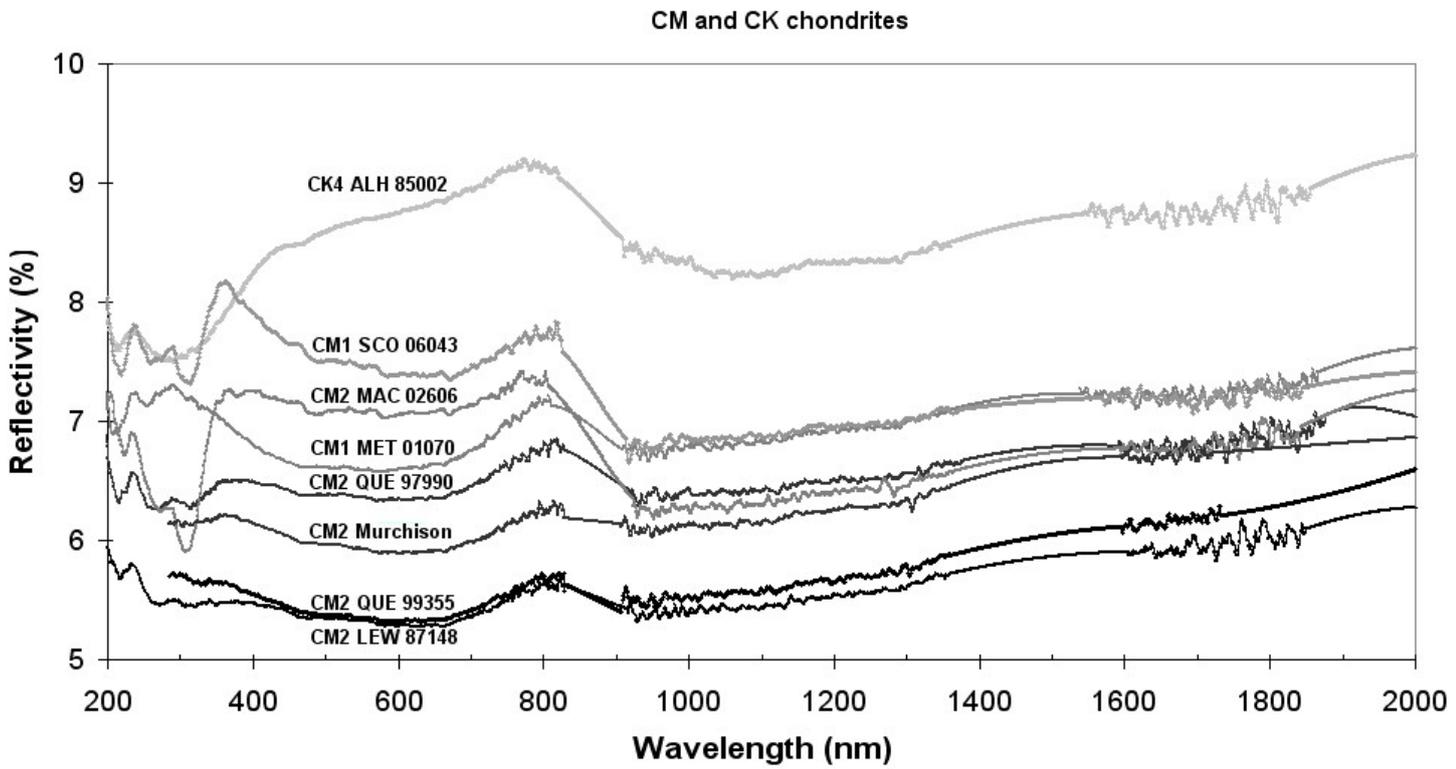

Figure 1a. UV-NIR spectra of CM and CK carbonaceous chondrites

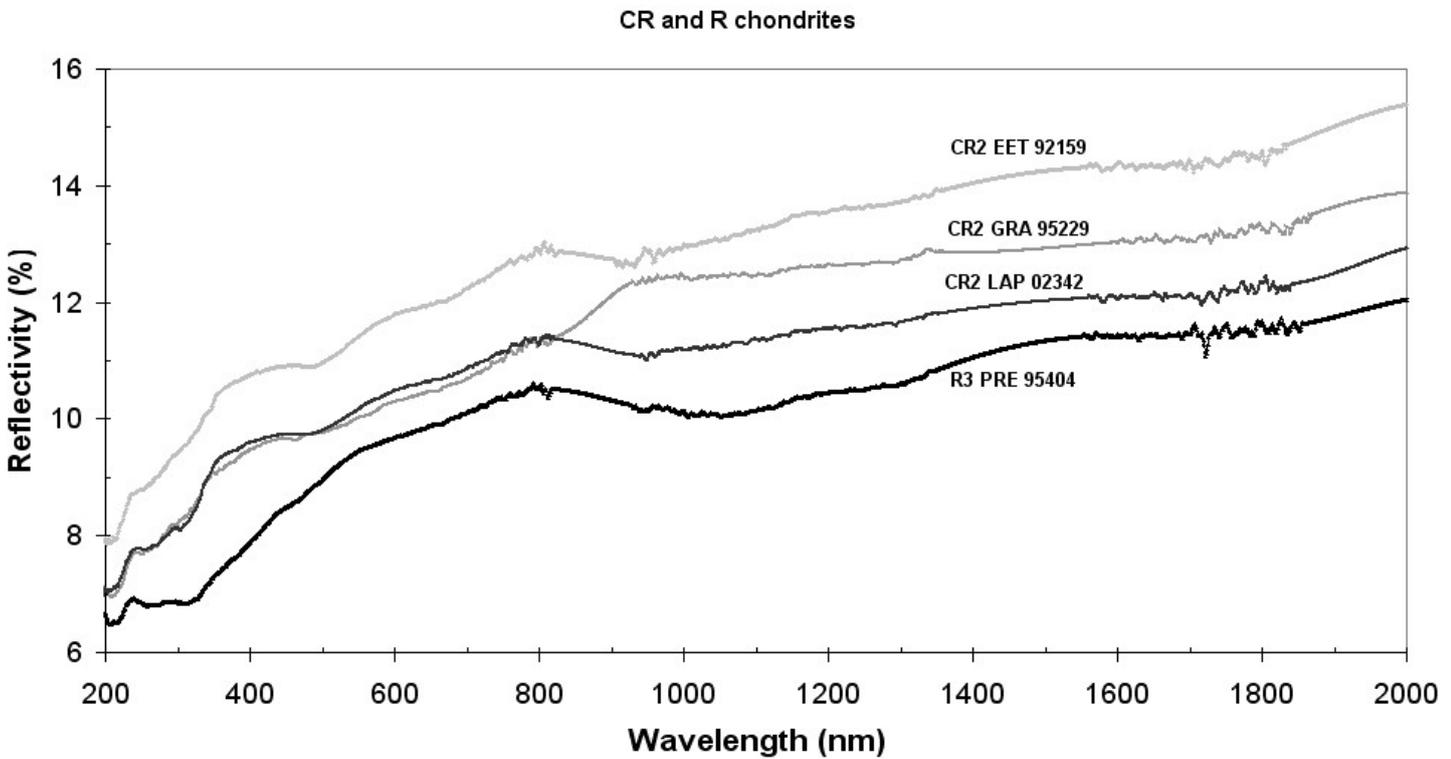

Figure 1b. UV-NIR spectra of CR and R carbonaceous chondrites

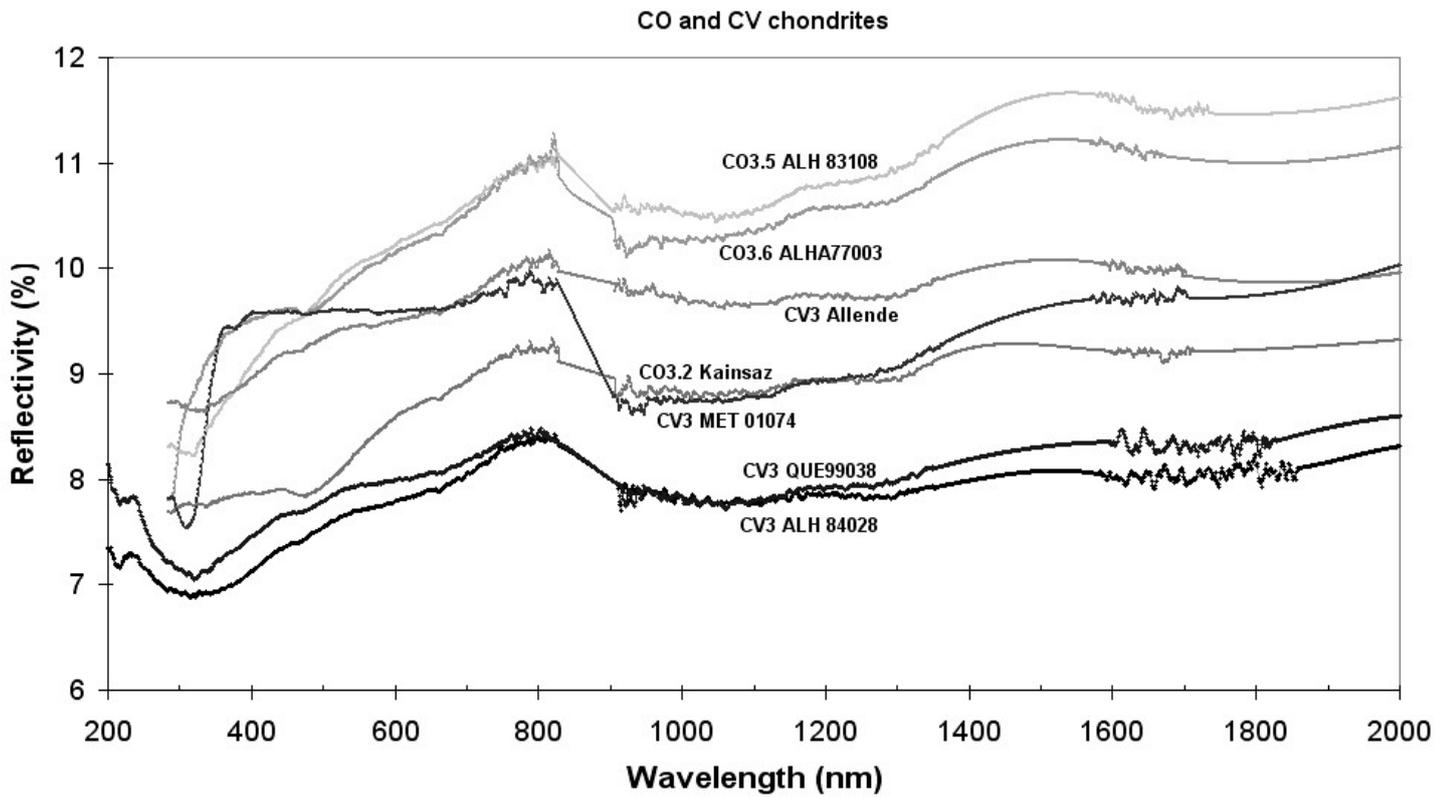

Figure 1c. UV-NIR spectra of CO and CV carbonaceous chondrites

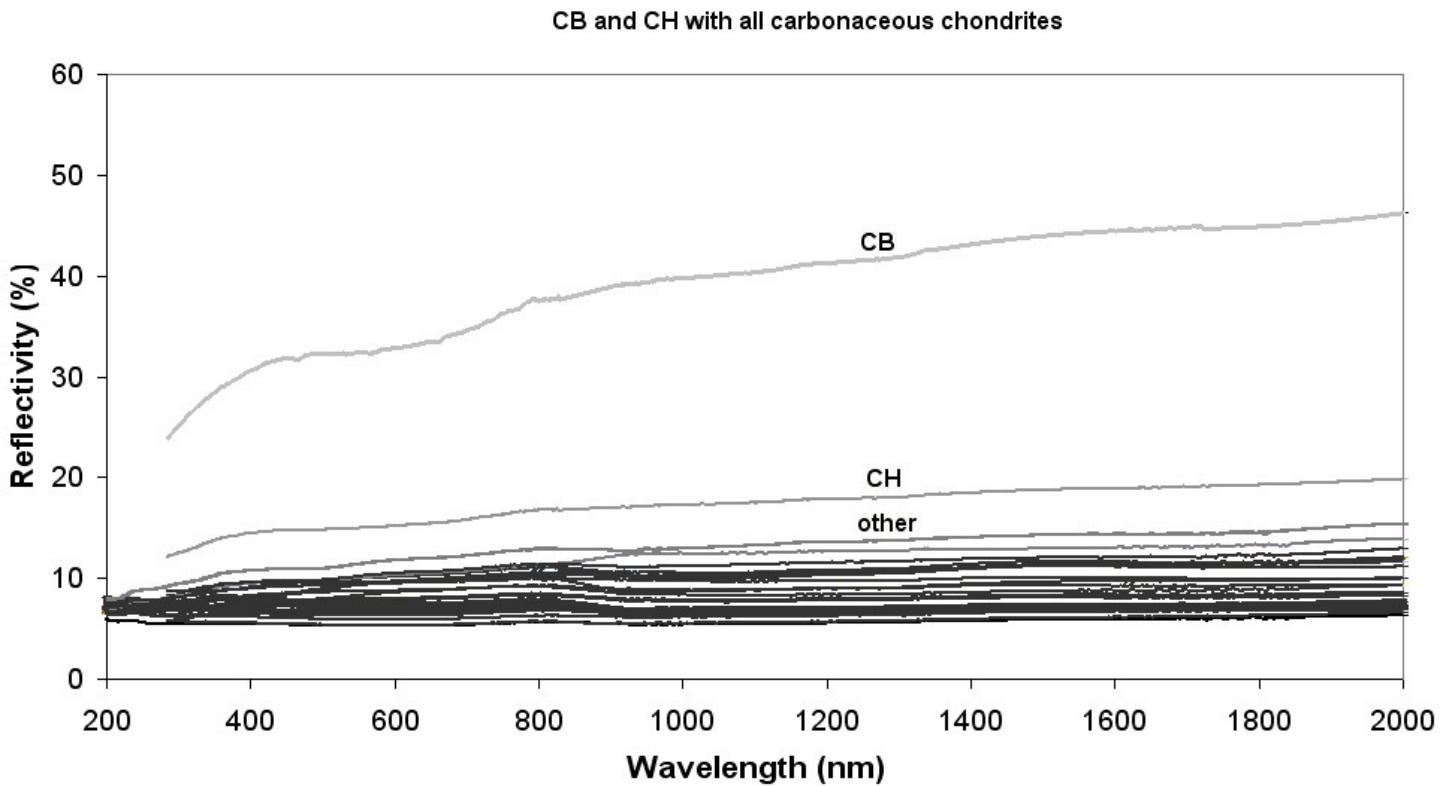

Figure 1d. UV-NIR spectra of all studied carbonaceous chondrite groups. Most of them have reflectivity between 5 and 15% along the full spectral window, but CH and CB chondrites.

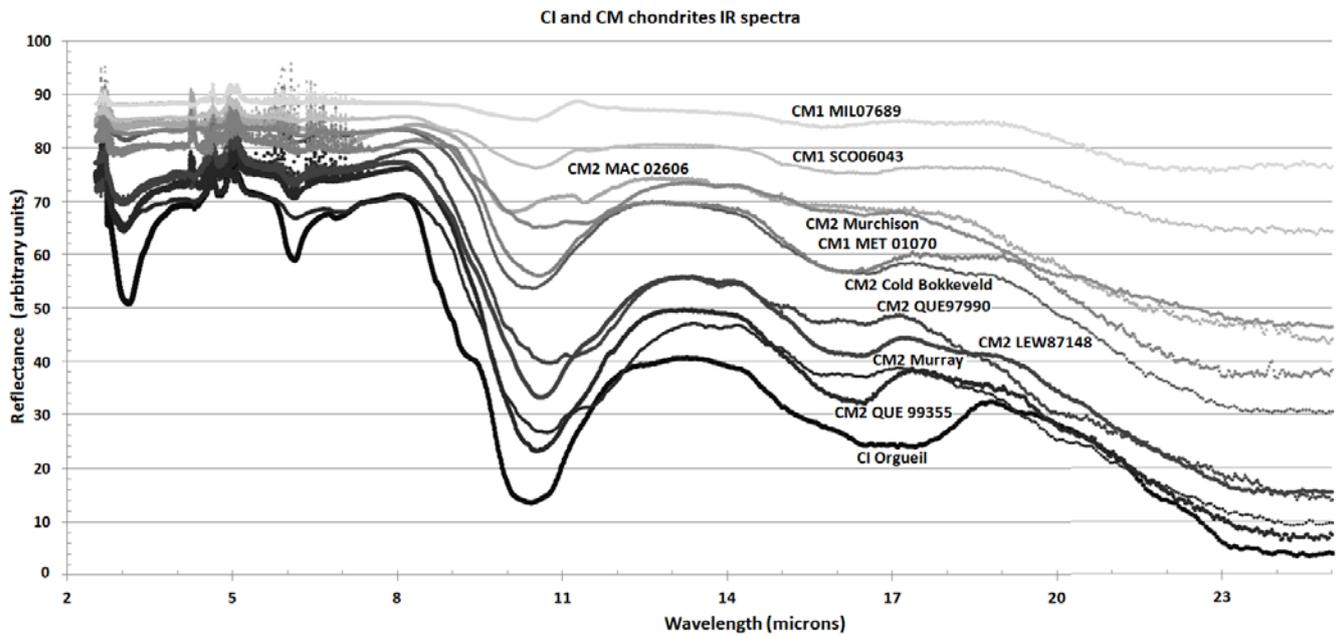

Figure 2a) IR spectra of CM carbonaceous chondrites from 2.5 to 25 µm. Notice the affinity with Orgueil CI chondrite, which is most distinguishable for showing deeper absorption bands than CM members.

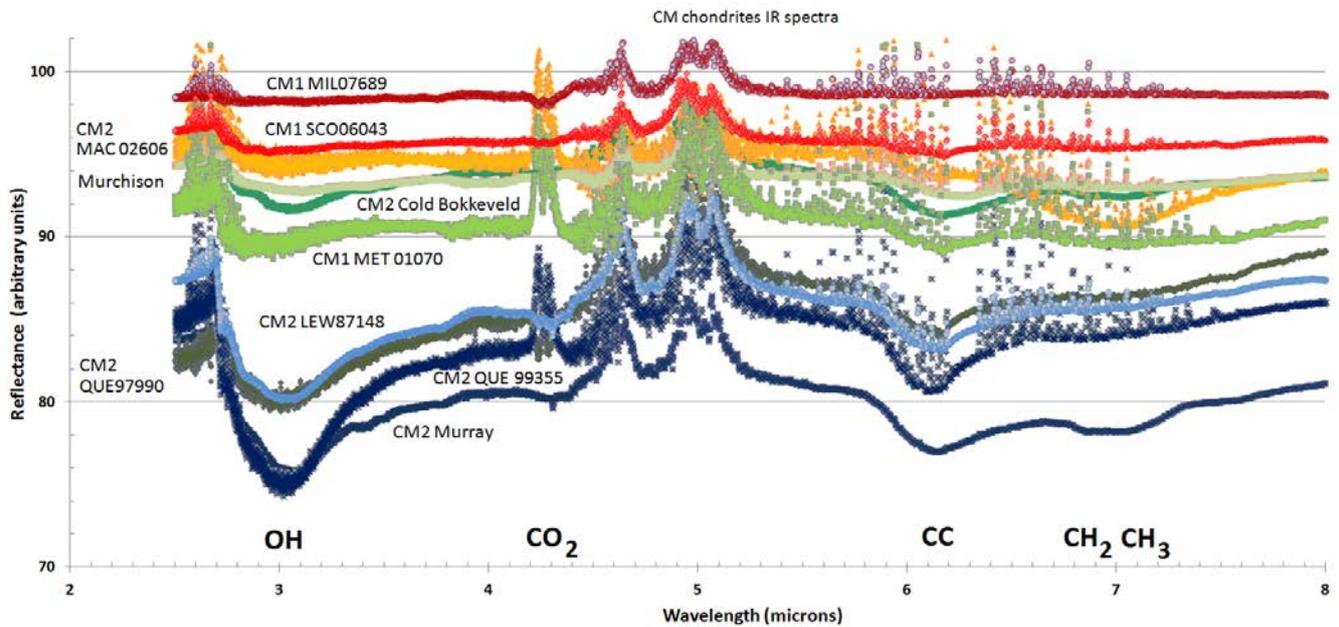

**Figure 2b (online only)** The 2.5 to 8 µm window to show the 3 µm band characteristic of the water bounded in phyllosilicates and other features described in Table 4. For example we marked the location of the prominent OH absorption band, the two emission peaks located around 4.25 µm from atmospheric $CO_2$, and the absorption bands of CC, $CH_2$, and $CH_3$.

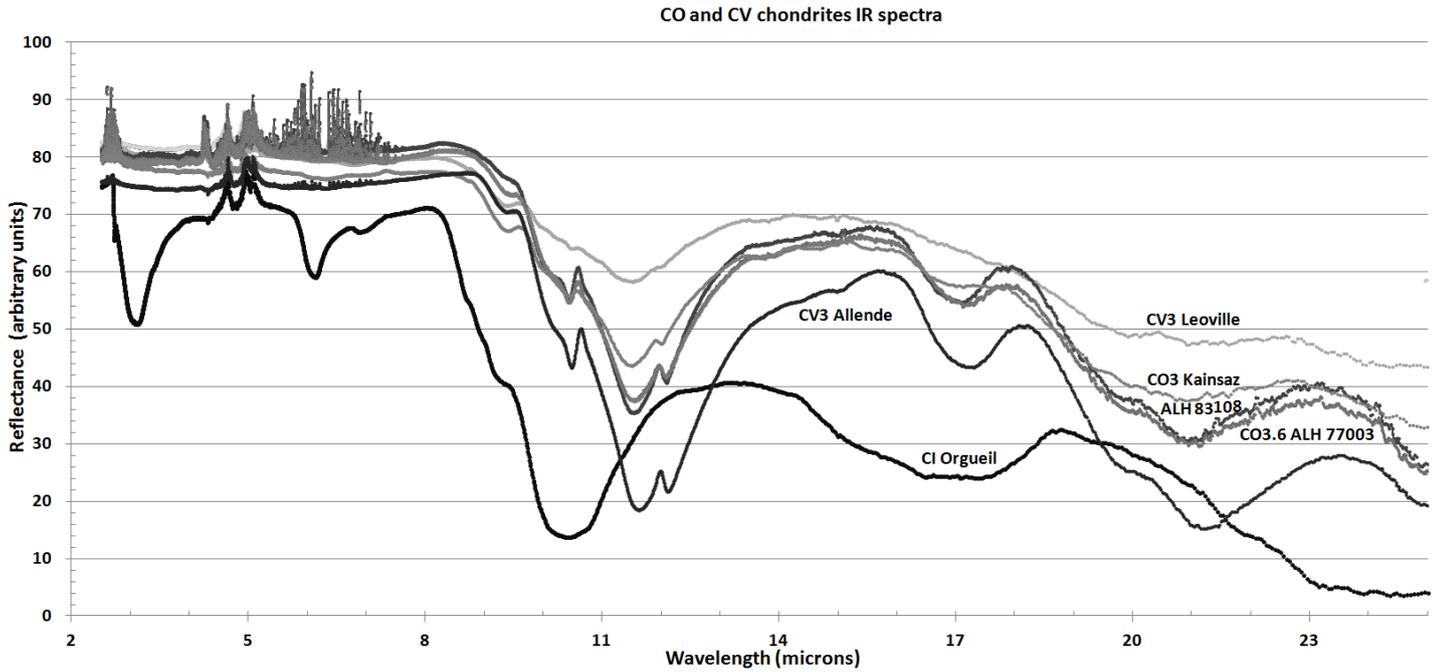

Figure 3. IR spectra of CO and CV chondrite groups in the 2.5 to 25 µm window. Notice the different location of the main absorption bands compared to Orgueil CI chondrite.

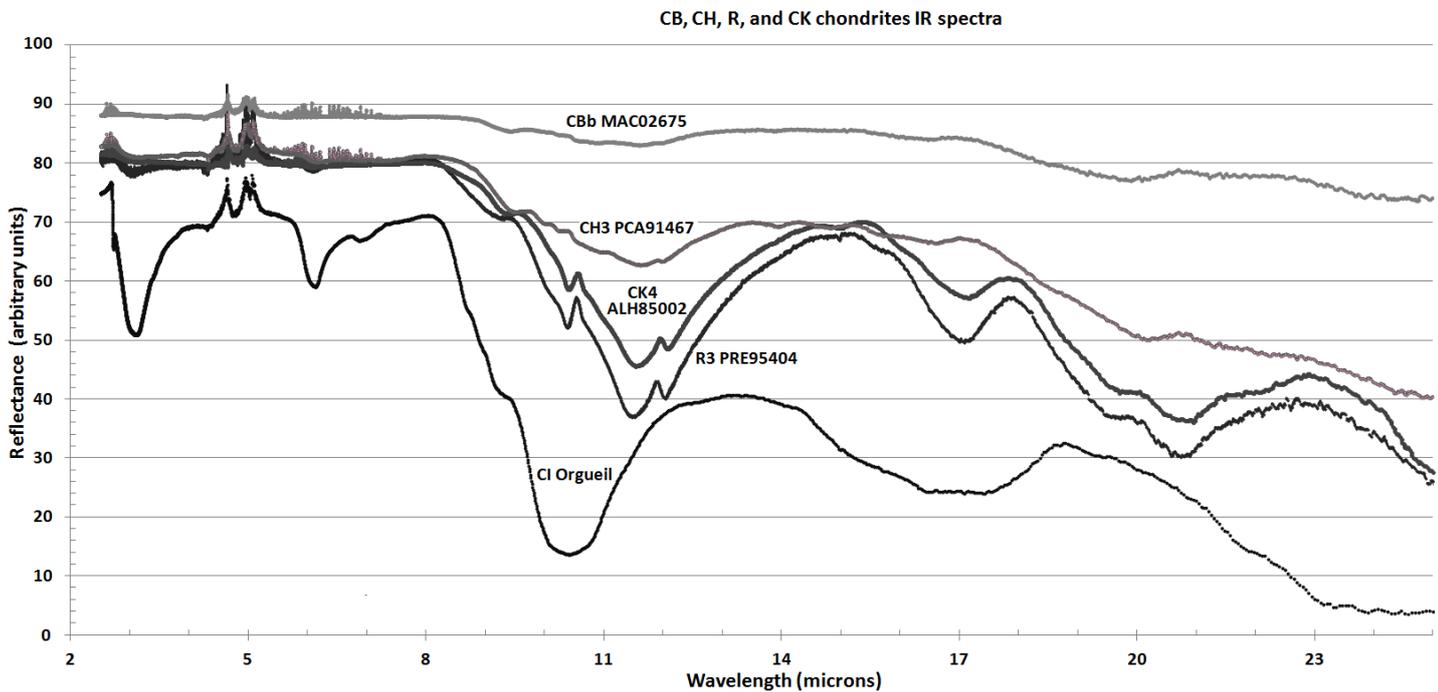

Figure 4. IR spectra of other carbonaceous chondrite groups in the 2.5 to 25 µm window. Notice the extreme reflectivity found for CB chondrite MAC 02675, not so surprising due to be about 60 vol% metal (Weisberg et al., 1988).

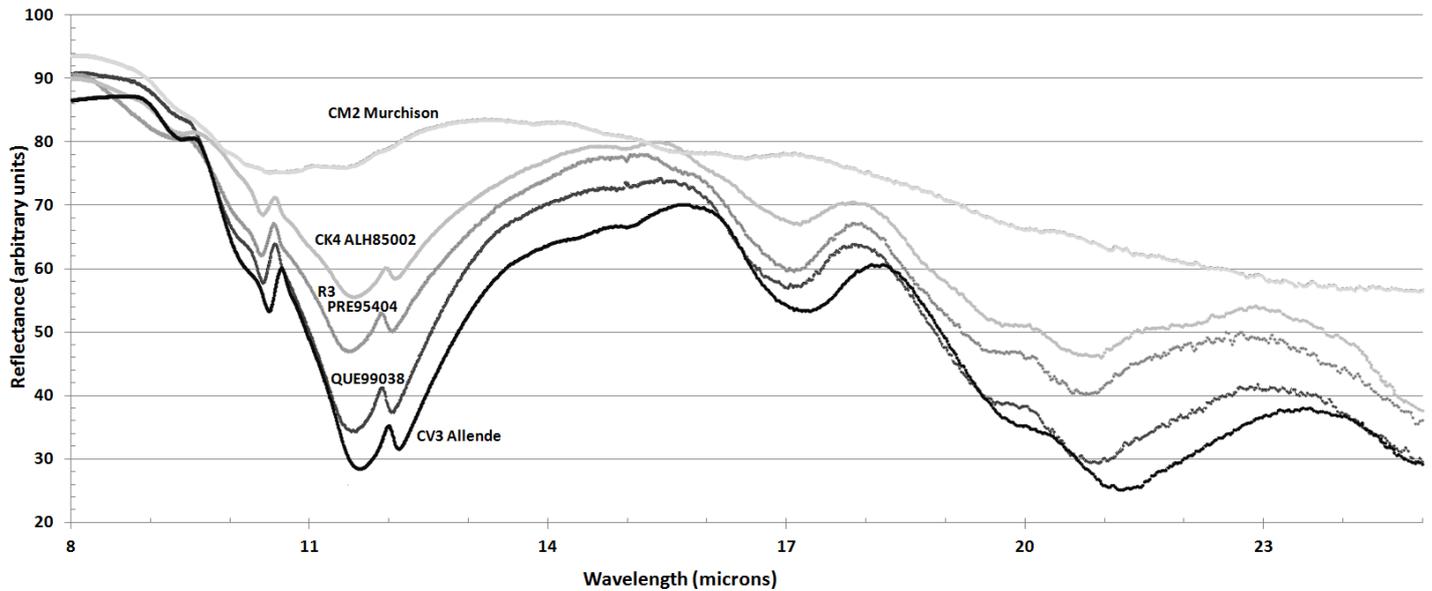

Figure 5. IR spectra of the currently classified CM chondrite QUE 99038 in the 8 to 25 µm window, compared with the spectra of the **CV3 Allende**, the R3 PRE 95404, the CK: ALH 85002, and the CM2 Murchison.

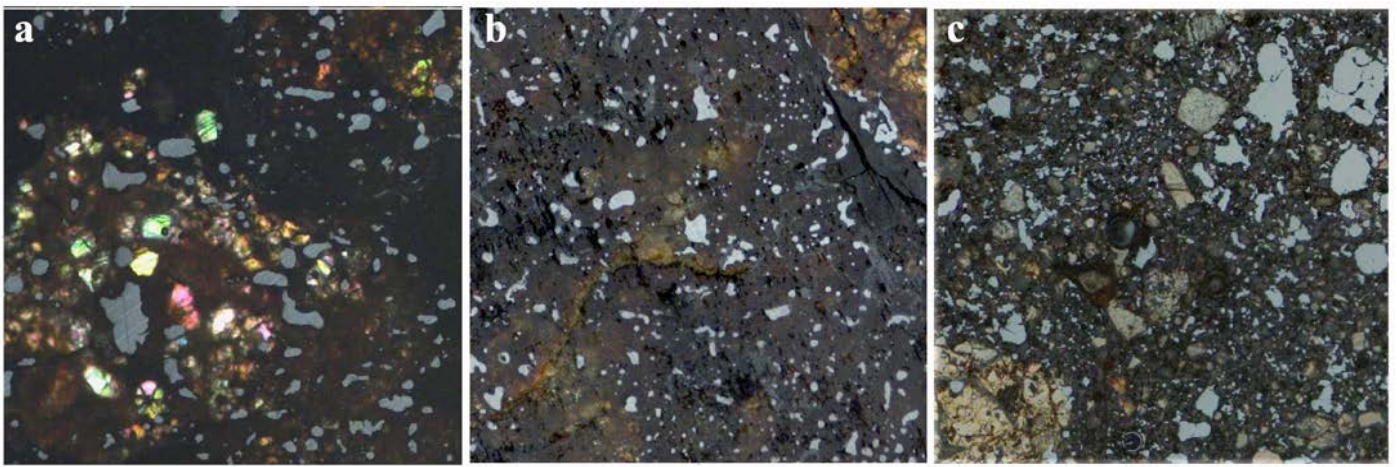

Figure 6. A 1 mm$^2$ window of three metal-rich carbonaceous chondrites discussed here exhibiting increasing amounts of metal from left to right. Reflectance images taken with a Zeiss petrographic microscope. The specimens are: a) CR2 chondrite GRA 95229, b) CR2 EET 92159, and c) CH3 PCA 91467.